\documentclass[twocolumn,aps,prl,amsmath,amssymb,superscriptaddress,longbibliography]{revtex4-2}
\usepackage[utf8]{inputenc}
\usepackage[T1]{fontenc}
\usepackage{lmodern}
\usepackage{nccmath}
\usepackage{bbm}

\usepackage{mathtools} 
\usepackage{hyperref} 
\usepackage{color}

\usepackage{multirow}

\usepackage[caption=false,labelformat=simple]{subfig}

\usepackage{tensor}
\usepackage{xcolor}
\def\AJ#1{{\color{black}{#1}\color{blue}}}
  
\begin{document}

\title{When Does Population Diversity Matter? A Unified Framework for Binary-Choice Dynamics} 

\author{Arkadiusz J\k{e}drzejewski}
\email{arkadiusz.jedrzejewski@ua.pt}
\affiliation{Department of Physics and I3N, University of Aveiro, 3810-193 Aveiro, Portugal}

\author{Jos\'{e} F. F. Mendes}
\affiliation{Department of Physics and I3N, University of Aveiro, 3810-193 Aveiro, Portugal}

\date{\today}
\begin{abstract}
We propose a modeling framework for binary-choice dynamics in which agents update their states using two mechanisms selected based on individual preference drawn from an arbitrary distribution. We compare annealed dynamics, where preferences change over time, and quenched dynamics, where they remain fixed. 
Our framework bridges gaps between existing models and provides a systematic approach to assess when individual-level diversity affects collective dynamics and when it can be effectively ignored.
We identify a constraint on transition probabilities that makes annealed and quenched dynamics equivalent. 
We show that when this condition is satisfied, the quenched dynamics reduces to a one-dimensional system, ruling out oscillatory behavior that may otherwise emerge.\\\\
Post-print of \href{https://doi.org/10.1103/4db2-7dpd}{Phys. Rev. Lett. \textbf{135}, 217401 (2025)}.\\
Copyright (2025) by the American Physical Society.

\end{abstract}

\maketitle\newpage

\textit{Introduction}---There is a broad class of binary-choice models where agents change their states following one of two mechanisms according to some probabilities.
In social and economic contexts, these mechanisms may correspond to different social responses, like conformity, nonconformity, or independence \cite{Jed:Szn:19}, different learning strategies (e.g., individual or social learning) \cite{Ren:etal:11,Ken:etal:18}, or heuristic rules (e.g., following the majority or minority  \cite{Oli:92}, or other threshold-based strategies  \cite{Vie:etal:20}).
They can also capture some economic effects such as network externalities (i.e., increasing livelihood of adoption as more agents choose one option) \cite{Dim:Gar:11,Dim:Gar:12} or the costs and the difficulties of adoption \cite{Byr:etal:16}.
Beyond sociophysics, the same model structure is relevant to other domains. The mechanisms may correspond to spins coupled to heat baths at different temperatures in statistical physics \cite{Gar:Lab:Mar:87,Tom:Oli:San:91,Tam:Ale:Gup:94}, to infection and recovery in epidemiology \cite{Kee:Eam:05}, or to selection and mutation in evolutionary theory \cite{Pag:Now:02}.
Depending on the context, the associated probabilities can be interpreted as individual preferences or relative frequencies of the mechanisms.
For consistency, we refer to them as preferences in this Letter.

\begin{figure}[!t]
	\centering
	\includegraphics[width=\linewidth]{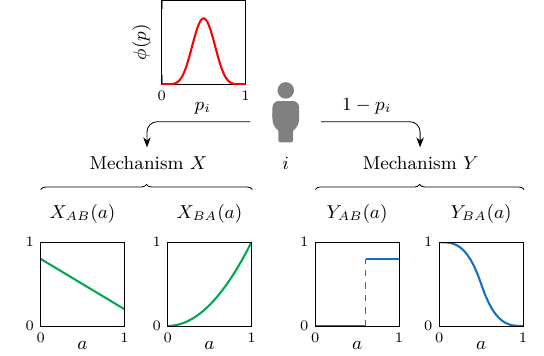}
	\caption{\label{fig:model-vis} Structure of the modeling framework. Agent $i$ must choose between two options, $A$ and $B$. The choice is made by using one of two mechanisms $X$ or $Y$. The mechanisms are defined by transition probabilities, which give the likelihood of changing the option from $A$ to $B$ ($AB$) and from $B$ to $A$ ($BA$).
    The transition probabilities can be arbitrary functions of the actual fraction of agents with option $A$ in the system, denoted by $a$. The functions shown are only examples.
    Agent $i$ has a personal preference towards mechanism $X$ represented by probability $p_i$ of choosing this mechanism. With complementary probability, $1-p_i$, mechanism $Y$ is chosen. The preference, $p_i$, is assigned to agent $i$ from the preference distribution $\phi(p)$, which is the same for all the agents.
    In the annealed case, preferences change in time, while in the quenched case, they stay fixed, as illustrated in Fig.~\ref{fig:que-ann-vis}.
 }
\end{figure}
In many of these models, we can find simplifying assumptions about the distribution of preferences and how these preferences evolve in time.
In general, each agent $i$ can have a personal preference $p_i$, which determines the probability of following one behavioral mechanism over the other: with probability $p_i$, the agent adopts one behavior, and with $1-p_i$, the alternative.
However, a common approach is to assign these preferences in an overly coarse manner, either assuming homogeneous population \cite{Oli:92,Vie:etal:20,Nyc:Szn:Cis:12,Nyc:Szn:13,Abr:Paw:Szn:19,Now:Szn:19,Bar:Gra:Szn:22,Now:Szn:20,Gar:Lab:Mar:87,Tom:Oli:San:91,Tam:Ale:Gup:94,Vie:Cro:16} or by dividing it into just two distinct groups \cite{Jed:Szn:17,Jed:Szn:22, Jav:Squ:15,Gra:Li:20,Vil:Mor:Sou:12,Kha:Tor:19,Tan:Mas:13,Vil:Sou:17,Kra:18}.
In the homogeneous case, all agents have the same fixed preference, $p_i=p$, which results in a one-point distribution of preferences in the system.
Some variants of the majority-vote model \cite{Oli:92}, the Galam model \cite{Gal:04,Bor:Gal:06,Sta:Mar:04}, the Sznajd model \cite{Lam:lop:Wio:05}, the $q$-voter model \cite{Nyc:Szn:Cis:12,Nyc:Szn:13,Abr:Paw:Szn:19}, or the threshold model  \cite{Now:Szn:19,Bar:Gra:Szn:22,Now:Szn:20} implement this approach.
For example, in the original majority-vote model \cite{Oli:92}, all agents share the same probability of following the majority, otherwise, they follow the minority. 
In the more heterogeneous case, the agents are divided into two subgroups with fixed characteristic behaviors.
This effectively creates a Bernoulli distribution of preferences: $p_i$ of each agent is set to either 0 or 1.
In this way, one group consistently follows one behavioral rule, whereas the other always follows the alternative.
These fixed behavioral types may correspond to roles such as conformists, contrarians \cite{Kha:Tor:19,Sta:Mar:04,Sch:04,Tan:Mas:13} or anticonformists \cite{Jed:Szn:17,Jed:Szn:22,Jav:Squ:15,Gra:Li:20}, independent agents \cite{Jed:Szn:17,Jed:Szn:22, Byr:etal:16,Szn:Szw:Wer:14}, or social and individual learners \cite{Yan:etal:21,Jed:Her:24,Wu:etal:25}.
For example, in some versions of the $q$-voter model \cite{Jed:Szn:17,Jed:Szn:22, Jav:Squ:15}, there is a fixed fraction of agents that are conformists ($p_i=0$), and the rest are nonconformists ($p_i=1$).
Although real-world populations may exhibit richer variation in behavioral preferences, only a few studies have explored more general distributions of preferences \cite{Yan:etal:21}, or explicitly pointed out that such extensions are possible \cite{Vil:Mor:Sou:12,Jed:Her:24}.
In this Letter, we go beyond these coarse assumptions, and we let each agent have an individual preference, $p_i$, drawn from an arbitrary distribution.

Another important modeling choice concerns whether individual preferences evolve or remain fixed over time.
In this regard, two distinct types of dynamics are commonly considered: annealed and quenched \cite{Byr:etal:16,Jed:Szn:20,Sta:Mar:04}.
In annealed dynamics, preferences fluctuate over time, capturing populations were behavioral inclinations are variable or context dependent.
In contrast, quenched dynamics fixes the values of preferences throughout the whole evolution of the system, modeling agents with stable traits.
This distinction is conceptually related to the person-situation debate in psychology, asking whether individual behaviors are better explained by personal traits or by situational factors \cite{Don:Luc:Fle:09}.
Some studies on agent-based modeling have drawn on this analogy, interpreting annealed disorder as reflecting the situation-oriented perspective, whereas quenched disorder as the personality-oriented one \cite{Byr:etal:16,Szn:Szw:Wer:14,Jed:Szn:20,Jed:Szn:17}.
For homogeneous populations, where all agents share the same preference, annealed and quenched dynamics are trivially identical.
For heterogeneous populations, however, the situation is less straightforward, and previous studies have reported mixed conclusions.
Some report that the macroscopic outcomes remain qualitatively similar \cite{Sta:Mar:04,Szn:Szw:Wer:14} or even the same \cite{Jed:Szn:17,Jed:Her:24}, whereas others find significant differences \cite{Jed:Szn:17,Jed:Szn:20, Byr:etal:16}.
Despite this, many studies choose one of these dynamics over the other, and the implications of this choice are not systematically compared.
Some works explicitly point out that examining how these two types of dynamics influence macroscopic outcomes is an interesting research direction \cite{Jed:Szn:19,Vil:Mor:Sou:12,Bar:Gra:Szn:22}.

In this Letter, we address these issues within one modeling framework that captures many previously studied binary-choice models.
Our framework allows us to systematically compare the behavior of models with arbitrary distribution of preferences under annealed and quenched dynamics, offering new insights into how preference heterogeneity and their temporal variability shape collective dynamics.
We identify the condition under which annealed and quenched dynamics lead to the same macroscopic outcome for any preference distribution, even heterogeneous ones.
In such cases, a heterogeneous population can be mapped into a homogeneous one, eliminating the effects of preference heterogeneity and ruling out oscillatory behavior.
This result not only clarifies apparent contradictions in previous studies but also simplifies the analysis of models that satisfy this condition. 
It also supports model design by indicating which models cannot account for cyclic processes.

\begin{table}[t]
\caption{\label{tab:prob} Probabilities that an agent in a given state keeps it or changes it to the alternative one under mechanisms $X$ and $Y$.
The fraction of agents in state $A$ in the system is denoted by $a$.
}
\centering
\renewcommand{\arraystretch}{1.5}
\begin{tabular}{c|cc|cc}
\toprule
\multirow{3}{*}{\parbox[t]{1.2cm}{Initial states}}                 & \multicolumn{4}{c}{\parbox[t]{6.2cm}{ Final states}}              \\
\cline{2-5}
               & \multicolumn{2}{c|}{Mechanism X}          & \multicolumn{2}{c}{Mechanism Y}                  \\ \cline{2-5}
 & \parbox[t]{1.55cm}{$A$}                        & \multicolumn{1}{c|}{\parbox[t]{1.55cm}{$B$}} & \parbox[t]{1.55cm}{$A$}                            & \multicolumn{1}{c}{\parbox[t]{1.55cm}{$B$}} \\\colrule
$A$                                   & \multicolumn{1}{c}{$1-X_{AB}(a)$} & \multicolumn{1}{c|}{$X_{AB}(a)$}                & \multicolumn{1}{c}{$1-Y_{AB}(a)$} & \multicolumn{1}{c}{$Y_{AB}(a)$}                \\
$B$                                   & \multicolumn{1}{c}{$X_{BA}(a)$} & \multicolumn{1}{c|}{$1-X_{BA}(a)$}                & \multicolumn{1}{c}{$Y_{BA}(a)$}     & \multicolumn{1}{c}{$1-Y_{BA}(a)$}               \\ \botrule
\end{tabular}
\end{table}

\textit{Modeling framework}---Consider a well-mixed population of agents, each of which can be in one of two states, $A$ or $B$, representing different opinions or choices. 
The state of an agent can change over time due to two distinct mechanisms, $X$ and $Y$, which govern the state updates. 
Each mechanism defines the transition probabilities for an agent to change from one state to the other.
These probabilities may, in general, be arbitrary functions of the current fraction of agents in state $A$, denoted by $a$, including polynomial (constant or higher order) \cite{Oli:92,Tan:Mas:13,Jed:Szn:17,Jed:Szn:22, Nyc:Szn:Cis:12,Nyc:Szn:13,Jav:Squ:15,Abr:Paw:Szn:19}, threshold \cite{Now:Szn:19,Bar:Gra:Szn:22,Now:Szn:20}, or more complex forms \cite{Gar:Lab:Mar:87,Tom:Oli:San:91,Tam:Ale:Gup:94,Pag:Now:02,Yan:etal:21, Wu:etal:25}, possibly with additional parameters \AJ{(see Supplemental Material~\cite{SM} for more examples)}.
For mechanism $X$, the probabilities of switching from $A$ to $B$ and from $B$ to $A$ are denoted by $X_{AB}(a)$ and $X_{BA}(a)$, respectively.
Analogously, for mechanism $Y$, we use $Y_{AB}(a)$ and $Y_{BA}(a)$.
The transition probabilities are summarized in Table~\ref{tab:prob}. 

At each update step, a randomly selected agent, labeled by $i$, changes its state according to mechanism $X$ with probability $p_i$ and according to mechanism $Y$ with complementary probability $1-p_i$. 
The parameter $p_i$ reflects an agent's individual preference for the mechanism $X$.
This preference is drawn independently for each agent from the preference distribution $\phi(p)$, which is the same for all the agents.
Figure~\ref{fig:model-vis} illustrates the defining elements of the model.

\begin{figure}[!t]
	\subfloat{\label{fig:que-ann-vis:a}}
    \subfloat{\label{fig:que-ann-vis:b}}
	\centering
	\includegraphics[width=\linewidth]{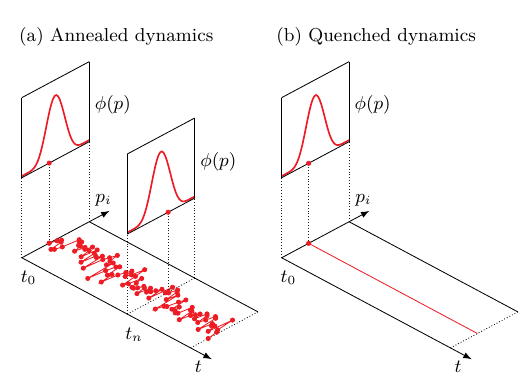}
	\caption{\label{fig:que-ann-vis} (a) Annealed dynamics: Agent $i$ is assigned a new preference towards mechanism $X$, $p_i(t)$, at each time step $t$ ($t_0$ and $t_n$ are highlighted in the plot as examples). This preference is drawn independently from the same distribution $\phi(p)$ for each agent and at each time step. 
    As a result, the preferences change in an uncorrelated way.
    (b) Quenched dynamics: At the beginning of the dynamics, $t_0$, agent $i$ is assigned a preference, $p_i(t_0)$, which stays fixed throughout time. 
    This preference is drawn independently from the same distribution $\phi(p)$ for each agent.
 }
\end{figure}

We distinguish between two dynamics, which correspond to different rates at which agents change their preferences \cite{Byr:etal:16}.
(i) Annealed dynamics: Agents change their preferences at each update step by drawing a new value of $p_i$ from the preference distribution $\phi(p)$, see Fig.~\ref{fig:que-ann-vis:a}.
This models highly adaptive behavior where preferences fluctuate over time.
(ii) Quenched dynamics: Agents do not change their preferences. The values of $p_i$ are drawn from the preference distribution $\phi(p)$ at the beginning of the process, and they remain unchanged, see Fig~\ref{fig:que-ann-vis:b}.   
This models population of agents persistent with their preferences, which can be associated to relatively stable personality traits.
Note that for both the dynamics, the empirical distribution of the preferences in the population of agents will approximate $\phi(p)$ as the number of agents becomes large.

\textit{Balancing condition}---Let us consider a model in which mechanism $X$ and mechanism $Y$ have balanced overall transition probabilities, i.e., 
\begin{equation}
\label{eq:balance}
     X_{AB}(a)+X_{BA}(a)=Y_{AB}(a)+Y_{BA}(a)
\end{equation}
for all $a\in[0,1]$. This spacial condition not only makes the quenched and annealed dynamics equivalent from a macroscopic point of view, but also ensures that the dynamics depends only on the mean of the preference distributions, $\bar{p}=\int p\phi(p)dp$, regardless of the shape or any other property of the distribution. 
Moreover, such systems cannot exhibit any oscillatory behavior.
To show it, let us analyze the differential equations governing the time evolution of the fraction of agents in state $A$ for both cases \AJ{(see Supplemental Material \cite{SM} for detailed derivation)}.

For the annealed dynamics, we have
\begin{equation}
\label{eq:rate-equation}
    \frac{da}{dt}=P_{BA}(1-a)-P_{AB}a,
\end{equation}
where $P_{BA}$ and $P_{AB}$ are the transition probabilities given by:
\begin{equation}
\label{eq:transition-rates-2}
\begin{split}
    P_{BA}&=\bar{p} X_{BA}(a)+(1-\bar{p})Y_{BA}(a),\\
    P_{AB}&= \bar{p} X_{AB}(a)+(1-\bar{p})Y_{AB}(a).
\end{split}
\end{equation}
From Eqs.~(\ref{eq:rate-equation}) and (\ref{eq:transition-rates-2}), we get that
\begin{equation}
\label{eq:time_evol_ann}
\begin{split}
    \frac{da}{dt}=&\bar{p}X_{BA}(a)+(1-\bar{p})Y_{BA}(a)-\left[Y_{BA}(a)+Y_{AB}(a)\right]a\\
    &-\left[X_{BA}(a)-Y_{BA}(a)+X_{AB}(a)-Y_{AB}(a)\right]\bar{p}a.
\end{split}
\end{equation}

For the quenched dynamics, let us divide the whole population into groups of agents with the same value of $p_i$.
The fraction of agents in the group with $p_i=p$ is given by $\phi(p)dp$.
We define $a_p$ as the fraction of individuals within each group that are in state $A$.
Consequently, the fraction of agents in state $A$ in the entire population is obtained by integrating $a_p$ over the distribution of preferences, $\phi(p)$, which gives:
\begin{equation}
\label{eq:apopulation}
    a=\int a_p\phi(p)dp.
\end{equation}
For each value of the probability $p$, we have the following equation for the time evolution of $a_p$:
\begin{equation}
    \label{eq:rate-que}
    \frac{da_p}{dt}=P^p_{BA}(1-a_p)-P^p_{AB}a_p,
\end{equation}
where $P^p_{BA}$ and $P^p_{AB}$ are the transition probabilities for the group of agents with  $p_i=p$:
\begin{equation}
\begin{split}
\label{eq:transition-rates-que}
    P^p_{BA}&=pX_{BA}(a)+(1-p)Y_{BA}(a),\\
    P^p_{AB}&= pX_{AB}(a)+(1-p)Y_{AB}(a).
\end{split}  
\end{equation}
To obtain the equation for the time evolution of $a$, we differentiate Eq.~\eqref{eq:apopulation} with respect to time. 
This gives
\begin{equation}
\label{eq:dadt_que}
\begin{split}
    \frac{da}{dt}=&\bar{p}X_{BA}(a)+(1-\bar{p})Y_{BA}(a)-\left[Y_{BA}(a)+Y_{AB}(a)\right]a\\
    &-\left[X_{BA}(a)-Y_{BA}(a)+X_{AB}(a)-Y_{AB}(a)\right]\\
    &\times\int pa_p\phi(p)dp.
\end{split}
\end{equation}
As seen, we cannot solve this equation without imposing the exact shape of the distribution $\phi(p)$.

The key point is that both the differential equations for annealed and quenched dynamics, i.e., Eqs.~\eqref{eq:time_evol_ann} and \eqref{eq:dadt_que}, simplify to 
\begin{equation}
\label{eq:da-balance}
\begin{split}
    \frac{da}{dt}=&\bar{p}X_{BA}(a)+(1-\bar{p})Y_{BA}(a)-\left[Y_{BA}(a)+Y_{AB}(a)\right]a.
\end{split}
\end{equation}
after applying the balancing condition given in Eq.~\eqref{eq:balance}. 
This formula depends only on the mean of the preference distribution, $\bar{p}$.
Although, for the annealed dynamics, the formula before simplification already was dependent only on $\bar{p}$, for the quenched dynamics, it was necessary to know the whole distribution, $\phi(p)$, to determine the time evolution of the system, as the last term of Eq.~\eqref{eq:dadt_que} was dependent on $\phi(p)$.

The differential equation describing the time evolution of $a$ under the annealed dynamics, Eq.~\eqref{eq:time_evol_ann}, forms a one-dimensional system. 
In such systems, oscillatory behavior cannot exist as spiral fixed points, closed orbits, and limit cycles are not possible \cite{Str:18}.
In contrast, the same model under the quenched dynamics generally leads to a higher-dimensional system, Eq.~\eqref{eq:rate-que}, where oscillations may arise.
However, if the balancing condition is satisfied, we have shown that we can reduce the system to a one-dimensional equation, Eq.~\eqref{eq:da-balance}.
Consequently, the oscillatory behavior is not possible in this case.

Finally, the fixed points of Eq.~\eqref{eq:da-balance}, denoted by $a^*$, are given by 
\begin{equation}
\label{eq:fixed-points-balance}
    \bar{p}=\frac{Y_{BA}(a^*)-a^*\left[Y_{BA}(a^*)+Y_{AB}(a^*)\right]}{Y_{BA}(a^*)-X_{BA}(a^*)},
\end{equation}
which implicitly defines the fixed points as a function of $\bar{p}$.
Thus, when the balancing condition holds, the fixed points depend only on the mean of the preference distribution, and they are the same for annealed and quenched dynamics.

\begin{figure*}[t!]
    \subfloat{\label{fig:fixed-points-dist:a}}
    \subfloat{\label{fig:fixed-points-dist:b}}
    \subfloat{\label{fig:fixed-points-dist:c}}
    \subfloat{\label{fig:fixed-points-dist:d}}
    \subfloat{\label{fig:fixed-points-dist:e}}
    \subfloat{\label{fig:fixed-points-dist:f}}
    \subfloat{\label{fig:fixed-points-dist:g}}
    \subfloat{\label{fig:fixed-points-dist:h}}
	\centering
	\includegraphics[width=\linewidth]{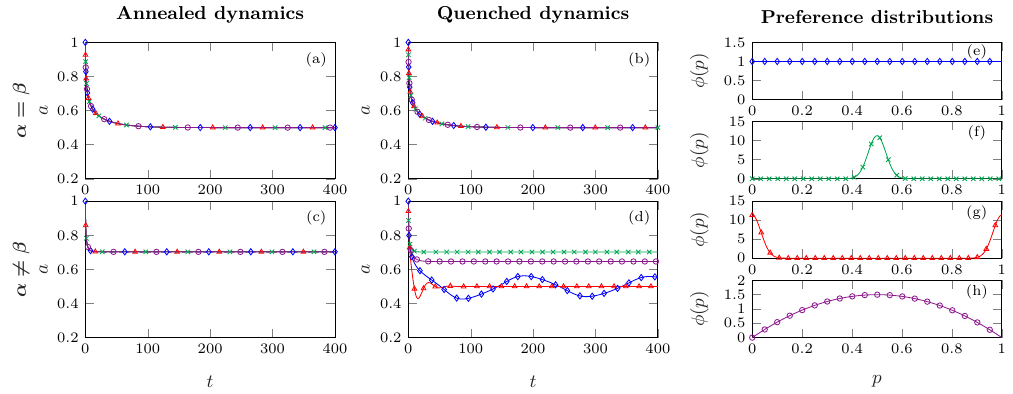}
	\caption{\label{fig:time-evol} Time evolution of the $q$-voter model with anticonformity \cite{Abr:Paw:Szn:19} for different distributions of preferences with the same mean: $\bar{p}=0.5$.
    Panels (a) and (b) correspond to the model parametrization that satisfies the balancing condition ($\alpha=\beta=6$), while  panels (c) and (d) show the case that does not satisfy the condition ($\alpha=6$, $\beta=2$).
    The four distributions of preferences are: (e) uniform distribution on [0, 1], (f) normal distribution centered at $p=0.5$ with variance $\sigma^2=1/800$, (g) mixture of two normal distributions centered at $p=0$ and $p=1$, both with variance $\sigma^2=1/800$, and (h) beta distribution with parameters: $\alpha_\text{B}=\beta_\text{B}=2$.
    Each distribution is represented by a unique color and symbol consistently across all panels.
    For the parametrization satisfying the balancing condition, the trajectories for (a) annealed and (b) quenched dynamics coincide for all distributions with the same mean, and any heterogeneous population can be mapped into a homogeneous one with that mean.
    When the balancing condition is not satisfied, only (c) annealed dynamics depends on the mean, while (d) quenched dynamics depends on the full distribution shape. 
    In this case, heterogeneous populations cannot be reduced to homogeneous ones, and they may exhibit oscillations.
 }
\end{figure*}
\textit{Result illustration}---Within our modeling framework, the $q$-voter model with anticonformity \cite{Abr:Paw:Szn:19} is defined by the following transition probabilities: $X_{BA}(a)=(1-a)^{\alpha}$, $X_{AB}(a)=a^{\alpha}$,  $Y_{BA}(a)=a^{\beta}$, and $Y_{AB}(a)=(1-a)^{\beta}$, where mechanisms $X$ and $Y$ are interpreted as anticonformity and conformity, respectively. 
The parameters $\alpha$ and $\beta$ determine the strength of the anticonforming and conforming interactions, whereas $p$ reflects the agents' preference towards anticonfomrity. 
We can see that the balancing condition is satisfied only when $\alpha=\beta$.

Figure~\ref{fig:time-evol} illustrates the time evolution of $a$ for this model under annealed and quenched dynamics for two cases: one satisfying the balancing condition [$\alpha=\beta$, shown in Figs.~\ref{fig:fixed-points-dist:a} and \ref{fig:fixed-points-dist:b}] and one that does not [$\alpha\neq\beta$, shown in Figs.~\ref{fig:fixed-points-dist:c} and \ref{fig:fixed-points-dist:d}].
The trajectories are presented for four different distributions of preferences that share the same mean, $\bar{p}=0.5$, see Figs.~\ref{fig:fixed-points-dist:e}-\ref{fig:fixed-points-dist:h}.
When the balancing condition holds, the trajectories for the annealed, Fig.~\ref{fig:fixed-points-dist:a}, and quenched dynamics, Fig.~\ref{fig:fixed-points-dist:b}, coincide for all distributions as only the mean preference matters.
In this case, any heterogeneous population can be mapped into a homogeneous one with a one-point distribution centered at the mean of the heterogeneous distribution.
This mapping rules out oscillations.
When the balancing condition is violated, the system behaves differently. 
For annealed dynamics, Fig.~\ref{fig:fixed-points-dist:c}, the trajectories still depend only on the mean.
However, for the quenched dynamics, Fig.~\ref{fig:fixed-points-dist:d}, the distribution shape drastically impacts the trajectory of the system, leading to monotonic convergence to different fixed points (purple $\circ$ and green $\times$), damped oscillatory convergence to a fixed point (red $\triangle$), or sustained oscillations related to a limit cycle (blue $\diamond$).
In this case, heterogeneous populations cannot be mapped into homogeneous ones, and they may exhibit oscillations under quenched dynamics.

\textit{Conclusions}---We introduced a general modeling framework for binary-choice dynamics in which agents update their states using one of two mechanisms selected based on their individual preference.
The mechanisms are defined by the transition probabilities, and they may represent different social behaviors, learning strategies, or decision rules depending on the model.
The preferences may change in time (annealed dynamics) or stay fixed (quenched dynamics), and they are drawn from a global preference distribution, allowing for modeling population diversity.
Such a framework captures a wide range of models previously studied in the literature and enables a systematic investigation on how the population diversity and the change of the rate of preference affect the population-level dynamics.

The key result of our analysis is the identification of the balancing condition---a constraint on the transition probabilities that has far-reaching consequences for the system behavior.
For the annealed dynamics, the time evolution of the system is described by a single differential equation that depends only on the mean of the preference distribution. 
The shape of the distribution is irrelevant, meaning that different distributions with the same mean lead to identical trajectories and fixed points, which represent the final state of the system.
Consequently, the system exhibits relatively simple behavior, and modeling the full distribution of preferences is unnecessary.
In contrast, for quenched dynamics, the time evolution is described by a set of coupled differential equations.
Solving them requires knowledge of the entire preference distribution, not just its mean.
Such a system allows for the emergence of richer dynamics, including oscillatory behavior.  
However, if the balancing condition is satisfied then: 
(1)~Annealed and quenched dynamics become equivalent, meaning that they lead to identical trajectories and fixed points. 
(2)~The trajectory and fixed points depend only on the mean of the preference distribution, making the shape of the distribution irrelevant. 
As in the annealed case, different distributions with the same mean lead to the same outcome.
Consequently, any heterogeneous population can be mapped into a homogeneous one with a one-point distribution at that mean.
(3)~Despite the high-dimensional nature of the system, the macroscopic dynamics can be described by a closed one-dimensional differential equation. This reduction implies that the system cannot exhibit any oscillatory behavior, including limit cycles, spiral fixed points, or damped oscillations.

Our results explain the origin of some previously reported mixed findings: in studies where annealed and quenched dynamics produced equivalent macroscopic behavior, the balancing condition was satisfied \cite{Jed:Szn:17,Jed:Her:24}; in those where differences emerged, it was not \cite{Jed:Szn:17,Jed:Her:24,Szn:Szw:Wer:14,Byr:etal:16}.
The results also clarify the findings of Ref.~\cite{Yan:etal:21}, where a model satisfying the condition appeared to produce different fixed-point diagrams for different preference distributions. However, these differences do not stem from the shapes of the distributions themselves, but rather from differences in their means.

The equivalence of the dynamics, together with the possibility of reducing a heterogeneous system to a homogeneous one when the balancing condition holds, not only  simplifies model analysis but also allows researchers to focus on the aspects of the model that truly drive its behavior.
Beyond explaining past results, the balancing condition also indicates promising research directions, as models that violate it are likely to exhibit different behaviors under alternative dynamics and preference distributions. 
In such cases, modeling of heterogeneity becomes meaningful in the quenched case, and the dynamics should be carefully selected.
Moreover, models that break the condition may exhibit oscillatory behavior, which makes them suitable for describing cyclic processes.
For instance, the homogeneous threshold model from Ref.~\cite{Bar:Gra:Szn:22}, studied only under annealed dynamics, satisfies the condition, so its quenched version will give the same results. 
In contrast, the $q$-voter model with anticonformity from Ref.~\cite{Abr:Paw:Szn:19}, in general, violates the condition and may yield different outcomes under quenched dynamics, justifying its further investigation.

Overall, the proposed framework with the balancing condition offers a structured way to examine the impact of agent-level heterogeneity on the system behavior.
However, one limitation of our analysis is the assumption of a well-mixed population.
While this assumption is common in early-stage modeling as it allows for analytical tractability, many models are eventually extended to structured populations represented by networks.
This can significantly affect the dynamics. 
Our framework does not account for these cases.
In particular, the differences between quenched and annealed dynamics may arise on networks even when the balancing condition holds in a well-mixed population, especially for sparse networks with small average node degree \cite{Jed:Szn:22}.
However, as the average node degree increases, the results tend to approach the well-mixed case. This suggests that for sufficiently dense networks the conclusions following from the balancing condition may still apply. Nevertheless,  the precise impact of the network structure may also depend on the specific model under consideration.
Extending the current analysis to the populations embedded in networks is therefore a relevant direction for future research.

\textit{Acknowledgments}---This work was developed within the scope of the project i3N, UID/50025 and LA/P/0037/2020, financed by national funds through the FCT/MEC. A. J. is supported by an i3N contract.

\textit{Data availability}---The data that support the findings of this article are openly available \cite{Jed:data}.

\bibliography{literature}
\newpage
\onecolumngrid
\section{End Matter}
\begin{figure*}[!b]
    \subfloat{\label{fig:fig-fixed-point-diagram:a}}
    \subfloat{\label{fig:fig-fixed-point-diagram:b}}
    \subfloat{\label{fig:fig-fixed-point-diagram:c}}
	\centering
	\includegraphics[width=0.95\linewidth]{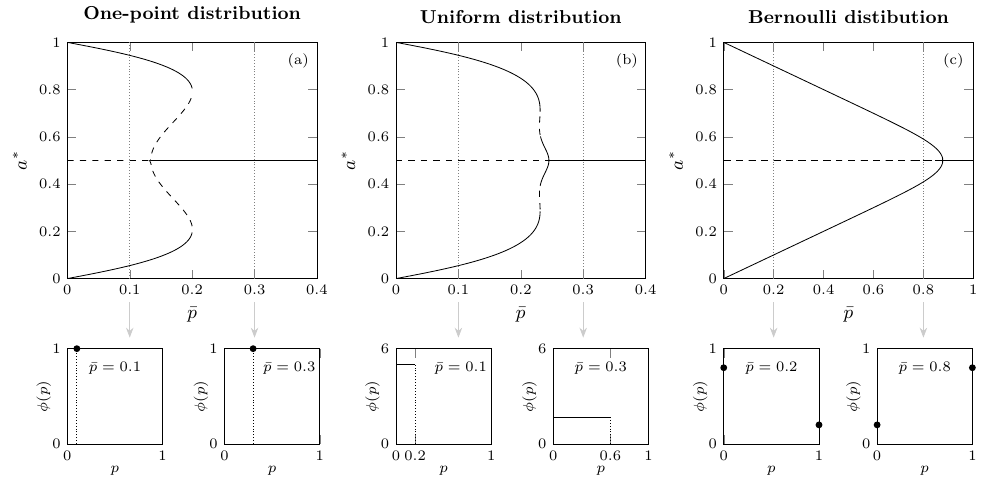}
	\caption{\label{fig:fig-fixed-point-diagram} Fixed-point diagrams under the quenched dynamics for the noisy threshold $q$-voter model \cite{Vie:etal:20} ($q=10$, $q_0=9$), which does not satisfy the balancing condition. Each panel corresponds to a different distribution of preferences: (a) sliding one-point distribution, (b) expanding uniform distribution, and (c) Bernoulli distribution.
    Each distribution leads to qualitatively different fixed-point diagram.
 }
\end{figure*}
\begin{figure*}[!t]
    \subfloat{\label{fig:fig-fixed-point-diagram-bc:a}}
    \subfloat{\label{fig:fig-fixed-point-diagram-bc:b}}
    \subfloat{\label{fig:fig-fixed-point-diagram-bc:c}}
	\centering
	\includegraphics[width=0.95\linewidth]{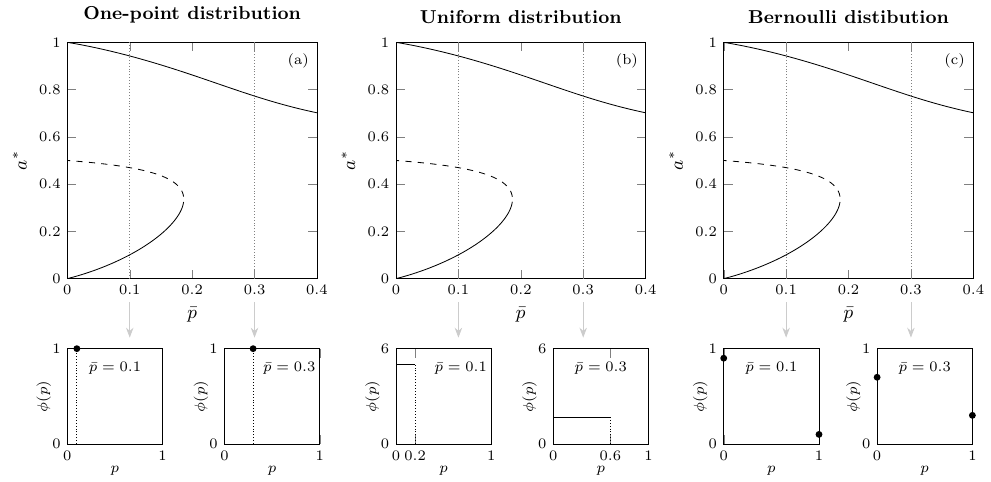}
	\caption{\label{fig:fig-fixed-point-diagram-bc} Fixed-point diagrams under the quenched dynamics for the dynamical system model of decision-making \cite{Yan:etal:21} ($q=1.5$ and $m=0.6$), which satisfies the balancing condition. Each panel corresponds to a different distribution of preferences: (a) sliding one-point distribution, (b) expanding uniform distribution, and (c) Bernoulli distribution.
    Despite differences in shape, all the distributions lead to the same fixed-point diagram.
 }
\end{figure*}

\textit{Result illustration}---The noisy threshold $q$-voter model \cite{Vie:etal:20} is defined by the following transition probabilities:  
\begin{equation}
            \begin{aligned}
                X_{BA}(a)=&1/2, \\
                X_{AB}(a)=&1/2, \\
                Y_{BA}(a)=&\sum_{k=q_0}^{q}\binom{q}{k}a^k(1-a)^{q-k}, \\
                Y_{AB}(a)=&\sum_{k=q_0}^{q}\binom{q}{k}(1-a)^ka^{q-k},
            \end{aligned}
            \begin{aligned}
            &\left.\vphantom{\begin{aligned}
                X_{BA}(a)=&1/2, \\
                X_{AB}(a)=&1/2, \\
            \end{aligned}}\right\rbrace\quad\text{Independence}\\
            &\left.\vphantom{\begin{aligned}
                Y_{BA}(a)=&\sum_{k=q_0}^{q}\binom{q}{k}a^k(1-a)^{q-k}, \\
                Y_{AB}(a)=&\sum_{k=q_0}^{q}\binom{q}{k}(1-a)^ka^{q-k},
              \end{aligned}}\right\rbrace\quad\text{Conformity}
            \end{aligned}
            \end{equation} 
where mechanism $X$ and $Y$ are interpreted as independence and conformity, respectively.
Under conformity, an agent adopts the opposite opinion if at least $q_0$ out of their $q$ randomly selected neighbors hold that opinion.
The parameter $p$ reflects the agents' preference towards independence.
This model does not satisfy the balancing condition.
Figure~\ref{fig:fig-fixed-point-diagram} presents fixed-point diagrams for three different distributions of preferences, from left to right: sliding one-point distribution localized at $\bar{p}$, extending uniform distribution supported on $[0,2\bar{p}]$, and Bernoulli distribution: $P(p=0)=1-\bar{p}$ and $P(p=1)=\bar{p}$. All these distributions are parametrized in a way that they have the same mean: $\bar{p}$.
Under the quenched dynamics when the balancing condition is not satisfied, the choice of the distribution can alter the number and the stability of fixed points.
For the one-point distribution, the system exhibits metastability, and it can switch abruptly between a state dominated by one option and a state with no clear majority, see Fig.~\ref{fig:fig-fixed-point-diagram-bc:a}.
For the uniform distribution, two transitions appear: first, a sudden switch between two distinct majority states, followed by a gradual shift towards a state where both options are equally likely, see Fig.~\ref{fig:fig-fixed-point-diagram-bc:b}. 
Finally, the Bernoulli distribution leads to a smooth transition, where the system gradually moves from a majority state to a state without it, see Fig.~\ref{fig:fig-fixed-point-diagram-bc:c}.

In contrast, all these distributions produce the same fixed-point diagram for a model that satisfies the balancing condition. As an example, let us consider the dynamical system model of decision-making \cite{Yan:etal:21}, which is defined by the following transition probabilities:  
\begin{equation}
            \begin{aligned}
                X_{BA}(a)=&m, \\
                X_{AB}(a)=&1-m, \\
                Y_{BA}(a)=&S(a), \\
                Y_{AB}(a)=&S(1-a),
            \end{aligned}
            \begin{aligned}
            &\left.\vphantom{\begin{aligned}
                X_{BA}(a)=&m, \\
                X_{AB}(a)=&1-m, \\
            \end{aligned}}\right\rbrace\quad\text{Individual learning}\\
            &\left.\vphantom{\begin{aligned}
                Y_{BA}(a)=&S(a), \\
                Y_{AB}(a)=&S(1-a),
              \end{aligned}}\right\rbrace\quad\text{Social learning}
            \end{aligned}
\end{equation}
where 
\begin{equation}
        S(x)=
        \begin{cases}
            \frac{1}{2}(2x)^q & \text{if }0 \leq x < 0.5,\\
            1-\frac{1}{2}\left[2(1-x)\right]^q & \text{if }0.5 \leq x \leq 1.
             \end{cases}
    \end{equation}
The mechanisms $X$ and $Y$ are interpreted as individual and social learning, respectively. 
The parameter $m$ is the relative merit of options, $q$ determines the type of conformity, whereas $p$ reflects the agents' preference towards individual learning. 
We can see that the balancing condition, Eq.~\eqref{eq:balance}, is always satisfied---as the transition probabilities for mechanisms $X$ and $Y$ sum up to 1 regardless of the values of $m$ and $q$.
Consequently, all the distributions lead to the same fixed-point diagram in Fig.~\ref{fig:fig-fixed-point-diagram-bc}.
Thus, the apparent differences in fixed-point diagrams reported in Ref.~\cite{Yan:etal:21} for different distributions do not arise from their shapes, but rather from the differences in their means.

The annealed and quenched dynamics are equivalent for the one-point distribution as it assigns the same preference to all the agents.
Therefore, panels~\ref{fig:fig-fixed-point-diagram:a} and \ref{fig:fig-fixed-point-diagram-bc:a} also effectively present the fixed-point diagrams for the corresponding models under the annealed dynamics.
With this in mind, Fig.~\ref{fig:fig-fixed-point-diagram} also illustrates that, in general, annealed and quenched dynamics lead to different results for the model that does not satisfy the balancing condition. 
In contrast, when this condition holds, the outcomes of annealed and quenched dynamics coincide for any preference distribution, as seen in Fig.~\ref{fig:fig-fixed-point-diagram-bc}, and any heterogeneous population can be mapped into a homogeneous one with the same mean preference.
 
\end{document}